\documentclass[11pt]{article}
\usepackage{slashed,cite}
\usepackage{latexsym}
\usepackage{amsmath}
\usepackage{amssymb}
\usepackage{lscape}
\usepackage{colordvi}
\usepackage{mathrsfs}

\csname @addtoreset\endcsname{equation}{section}
\newcommand{\ft}[2]{{\textstyle\frac{#1}{#2}}}
\def\rmi{{\,\rm i\,}}
\newsavebox{\uuunit}
\sbox{\uuunit}
    {\setlength{\unitlength}{0.825em}
     \begin{picture}(0.6,0.7)
        \thinlines
        \put(0,0){\line(1,0){0.5}}
        \put(0.15,0){\line(0,1){0.7}}
        \put(0.35,0){\line(0,1){0.8}}
       \multiput(0.3,0.8)(-0.04,-0.02){12}{\rule{0.5pt}{0.5pt}}
     \end {picture}}

\def\rmi{{\,\rm i\,}}
\begin{document}
\setcounter{page}{0}
\thispagestyle{empty}
\begin{flushright}
KUL-TF/03-20\\
hep-th/0310181\\[4cm]
\end{flushright}
\begin{center}
{\Large \textsc{A Generalisation of (Very) Special Geometry}}\\[3cm]
{\bf Jos Gheerardyn}\footnote{Aspirant FWO}\\
\vspace{1cm}
\textit{Instituut voor Theoretische Fysica, Katholieke Universiteit Leuven \break
Celestijnenlaan 200D, B-3001 Leuven, Belgium}\\[.5cm]
\texttt{Email : jos.gheerardyn@fys.kuleuven.ac.be}\\[4cm]
{\bf Abstract}\\[.5cm]
\begin{quote}
{\small We construct non-Abelian $\mathcal{N}=2$ on-shell vector multiplets in five and in four dimensions. Closing of the supersymmetry algebra imposes dynamical constraints on the fields, and these constraints should be interpreted as equations of motion. If these field equations should not be derivable from an action, we find that supersymmetry allows a broader class of target-space geometries than the familiar rigid (very) special manifolds. These theories moreover have more general potentials due to the possibility of including Fayet-Iliopoulos terms in the non-Abelian case. We show that by introducing an action, we recover the standard results. Finally, we relate the five- and the four-dimensional theories through dimensional reduction and discuss the corresponding generalised r-map.}
\end{quote}
\end{center}
\newpage
\section{Introduction}
During the past decade, special geometry has proven to be a rich and interesting subject. One of the main reasons why these geometries got so much attention is that this class of manifolds is quite restrictive and hence manageable. 

Theories with 32 supersymmetries are completely fixed, and the only possible set of fields is the supergravity multiplet. In theories with 16 supersymmetries there are already matter multiplets, but only a discrete set of possible target manifolds exists. Subtleties aside, these theories are completely fixed by the number and type of the  multiplets considered. The next possible theories have eight supersymmetries and for those, there is already enough freedom to define the target space geometry through continuous families of coupling functions. Because of the large amount of symmetry though, a lot of structure is imposed on these coupling functions. Therefore, the set of special manifolds is restrictive.

Rigid special geometry was first studied in \cite{Sierra:1983cc,Gates:1984py}. One of the most important applications is in the research on dualities, initiated by the Seiberg-Witten papers \cite{Seiberg:1994rs,Seiberg:1994aj}. Other important connections with special geometry were found in the study of Calabi-Yau compactifications \cite{Seiberg:1988pf,Ferrara:1989vp} and in the AdS/CFT correspondence \cite{Aharony:1999ti}. More recently, with the advent of brane-world scenarios \cite{Randall:1999ee,Randall:1999vf}, $\mathcal{N}=2$ supergravity again captioned attention. 

Because of this reconsideration of $\mathcal{N}=2$ supergravity, the target space geometry of the corresponding rigid supersymmetric theories is also studied. For instance, in constructing supergravity theories through the  superconformal tensor calculus approach  \cite{VanProeyen:2001wr,Bergshoeff:2001hc,Bergshoeff:2002qk}, one starts with superconformal matter multiplets for which the superconformal gauge fields act as a background. In these cases, the target space geometries are those of the rigid theories, possibly with some extra structure.\footnote{E.g. a conformal Killing vector field.} 

In \cite{Bergshoeff:2002qk} it was found that the target space for rigid hypermultiplets can be a hypercomplex space when the equations of motion are not derivable from an action. If they are, there is a compatible metric and the hypercomplex manifold becomes hyperk\"ahler. When such a theory is coupled to vector multiplets by gauging isometries in the hypercomplex target space, one encounters difficulties as the vector multiplets are off-shell, and dynamically coupling a theory with an action to one without, does not seem to make much sense. This is the main motivation of the present letter. We will construct an on-shell vector multiplet and will find that the possibilities for the target manifold are more general than in the usual off-shell case. 

Theories defined by equations of motion that are not derivable from an action are quite frequently encountered. There are different reasons why an action principle might not be at our disposal. First of all, the theory can contain an (anti-)self-dual field strength, like in the case of type IIB supergravity. In these cases, at least a simple action formulation does not exist. Another possibility is that the theory is constructed using generalised Scherk-Schwarz reduction with a symmetry that leaves the equations of motion invariant, but not the action. The reduction should then be performed on the equations of motion, and the resulting equations might not be derivable from an action any more. A quite well-known example is the massive IIA supergravity constructed using the scaling symmetry of the equations of motion of eleven dimensional supergravity \cite{Howe:1998qt,Lavrinenko:1998qa,Gheerardyn:2002wp} but other examples exist \cite{Hull:2003kr}. A final class of classical supersymmetric field theories for which there is no action are the ones with a non-Riemannian target space and a torsionless affine connection on the tangent bundle. The above mentioned hypercomplex manifolds for rigid hypermultiplets are examples of the latter. Other examples are quaternionic spaces in $\mathcal{N}=2$ supergravity coupled to hypermultiplets \cite{Bergshoeff:2001hc} and complex flat geometries in the $\mathcal{N}=1$ four-dimensional Wess-Zumino multiplet \cite{Stelle:2003rr}. We close this paragraph by mentioning that in string theory, one can find the equations of motion of the background fields directly by demanding the vanishing of the beta functions, to ensure conformal invariance. Afterwards, an action might (or might not) be constructed.

The content of this letter is as follows. In Section \ref{5d} the rigid five-dimensional on-shell vector multiplet will be constructed and the defining conditions for the geometry will be found. We will also include Fayet-Iliopoulos terms in our discussion. In Section \ref{action} we will use the Batalin-Vilkovisky formalism to show that by demanding the existence of an action, the target space becomes a very special real manifold. In Section \ref{4d} we will repeat the construction in four dimensions and compare our results to the well-known off-shell theories with special K\"ahler target manifolds in Section \ref{4daction}. In Section \ref{dimred} we will link the results of the previous Sections by dimensional reduction. Our final Section \ref{conclusions} contains the conclusions. An Appendix with conventions is also added. 
\section{On-Shell Vector Multiplet in Five Dimensions}\label{5d}
The purpose of this Section is to construct an on-shell $\mathcal{N}=2$ vector multiplet in five dimensions.
The fields should form a representation of the super-Poincar\'e group and should be charged under the action of the gauge group. The superalgebra contains an $\mathfrak{su}(2)$ R-symmetry, under which the gaugino is charged. The fields composing the multiplet, together with their numbers of degrees of freedom and their $\mathfrak{su}(2)$ weights are given in Table \ref{fieldsdof}, and as is well-known, the number of bosonic and fermionic degrees of freedom should match.\footnote{I.e. after elimination of degrees of freedom with the gauge invariances and the equations of motion of the system.}
\begin{table}[!htb]
\begin{center}
\begin{tabular}{|c|c|c|}
\hline
Field&$\mathfrak{su}(2)$&on-shell dof\\
\hline \hline
$\sigma$&1&1\\
$A_\mu$&1&3\\
$\psi^{i}$&2&4\\
\hline
\end{tabular}
\end{center}
\caption{Fields in the $\mathcal{N}=2, d=5$ vector multiplet}\label{fieldsdof}
\end{table}
If we would take the fields off-shell, the number of fermionic degrees of freedom would double, while the corresponding number of bosonic modes would only increase by one. Therefore, to compensate this mismatch between bosons and fermions, the off-shell vector multiplet contains an additional auxiliary bosonic field in the $\mathbf{3}$ of $\mathfrak{su}(2)$. 
 
The commutator of two supersymmetries should always contain a translation, but in the most general off-shell case, it turns out that the algebra also contains a field-dependent gauge transformation \cite{Gunaydin:1984bi,Bergshoeff:2002qk}.
\begin{equation}\label{algebra}
[\delta_1,\delta_2]=\delta_P(\ft12\bar{\epsilon}_2\gamma^\mu\epsilon_1)+\delta_G(-\ft12\rmi \sigma^I \bar{\epsilon}_2\epsilon_1)\; ,
\end{equation}
where $\delta_P$ is the translation and $\delta_G$ the gauge transformation. This algebra closes on all fields. One then proceeds by building a supersymmetric action, which is completely specified by a symmetric, gauge-invariant three-tensor $C_{IJK}$ (see also Section \ref{action}). The target space geometry is then called rigid (or affine) very special real and is Riemannian with metric $C_{IJK}\sigma^K$.

In on-shell multiplets, the closing of the supersymmetry algebra imposes dynamical constraints on the fields. More specifically, calculating the commutator of two supersymmetries on the fermions, one finds that there appears a non-closure functional of the fields at the right-hand side. This functional is then interpreted as the equation of motion for the fermions. One can then find the other equations of motion by considering the supersymmetry transformation of the latter. 

We start with the following transformation rules for the fields.
\begin{eqnarray}
\delta \sigma^I&=&\ft12 \rmi \bar{\epsilon}\psi^I-gf_{JK}{}^I\alpha^J\sigma^K\; ,\label{transfsigma2} \\
\delta A_\mu^I&=&\ft12 \bar{\epsilon}\gamma_\mu \psi^I+\partial_\mu\alpha^I+gf_{JK}{}^IA_\mu^J\alpha^K\; , \label{transfA2}\\
\delta \psi^{iI}&=&-\ft12 \rmi \slashed{\mathfrak{D}}\sigma^I\epsilon^i-\ft14\slashed{F}^I\epsilon^i+A^{(ij)I}\epsilon_j -gf_{JK}{}^I\alpha^J\psi^{iK}\; , \label{transfpsi2} 
\end{eqnarray}
where $\epsilon^i$ is the parameter for supersymmetry transformations and $\alpha^I$ the one for gauge transformations.
We found these rules by taking arbitrary transformations for the vector and the gaugino, and asking compatibility with the supersymmetry algebra. In this process, it turned out that in the most general case, the supersymmetry algebra is (\ref{algebra}) and the fields should transform as mentioned above. Another important point to make is that the field-dependent (and yet unknown) object $A^{(ij)I}$ in (\ref{transfpsi2}) (which transforms in the $\mathbf{3}$ of the R-symmetry group) stands in place of the auxiliary field in the off-shell case.

The field strength $F^I_{\mu \nu}$, the gauge-covariant derivative $\mathfrak{D}_\mu$ on a general object $Y^I$ transforming in the adjoint of the gauge group and the gauge covariant d'Alembertian are defined as
\begin{eqnarray}
F^I_{\mu \nu}&=&2\partial_{[\mu}A_{\nu]}^I+gf_{JK}{}^IA_\mu^JA_\nu^K\; , \nonumber\\
\mathfrak{D}_\mu Y^I&=&\partial_\mu Y^I+gf_{JK}{}^IA_\mu^JY^K\nonumber\\
\Box \sigma^I&=&\partial_\mu \mathfrak{D}^\mu \sigma^I+gf_{JK}{}^IA_\mu^J\mathfrak{D}^\mu \sigma^K\; .
\end{eqnarray}
It is now easy to check that the algebra (\ref{algebra}) is realised on the bosons. To do the same calculation for the gaugino, we need an Ansatz for the transformation of the field-dependent object $A^{(ij)I}$, which we take to be 
\begin{equation}\label{transfAijt}
\delta_Q A^{(ij)I}(\sigma,A_\mu,\psi^i)=-\bar{\epsilon}_k\zeta^{k,(ij)I}\; .
\end{equation}
where $\zeta^{k,(ij)I}$ is a field-dependent spinor in the $\mathbf{2}\times \mathbf{3}$ of $\mathfrak{su}(2)$. Expanding $\zeta^{k,(ij)I}=\varepsilon^{k(i}\zeta^{j)I}+\zeta^{(ijk)I}$ we can see that $\zeta^{(ijk)I}$ should be zero by trying to close the algebra (\ref{algebra}) on the fermions. In conclusion, the transformation rule (\ref{transfAijt}) becomes
\begin{equation}\label{transfAij}
\delta_Q A^{(ij)I}(\sigma,A_\mu,\psi^i)=\bar{\epsilon}^{(i}\zeta^{j)I}\; .
\end{equation}
The algebra on the fermions $\psi^{iI}$ then yields
\begin{eqnarray}
[\delta_1,\delta_2]\psi^{iI}&=&\delta_P(\ft12\bar{\epsilon}_2\gamma^\mu\epsilon_1)\psi^{iI}+\delta_G(-\ft12 \sigma^I \bar{\epsilon}_2\epsilon_1)\psi^{iI}\nonumber\\&&-\ft3{16}\bar{\epsilon}_2\epsilon_1\Gamma^{iI}-\ft3{16}\bar{\epsilon}_2\gamma^\mu \epsilon_1\gamma_\mu\Gamma^{iI} -\ft1{16}\bar{\epsilon}_2^{(i}\gamma^{\mu \nu}\epsilon_1^{j)}\gamma_{\mu \nu}\Gamma_j^I\; ,\\
\Gamma^{iI}&=&\slashed{\mathfrak{D}}\psi^{iI}+\rmi gf_{JK}{}^I\sigma^J\psi^{iK}-2\zeta^{iI}\; ,
\end{eqnarray}
where the non-closure functional $\Gamma^{iI}$ is the equation of motion for the fermions.

To know the explicit expression for $\zeta^{iI}$ we need the field-dependence of the object $A^{(ij)I}$, which can be inferred from its transformation rule (\ref{transfAij}) together with dimensional considerations. Starting from the dimensions of the fields shown in Table \ref{dimtable} and supposing that there do not exist objects of negative dimension, the most general expression for $A^{(ij)I}$ reads
\begin{equation}\label{AnsatzA}
A^{(ij)I}=gf^{(ij)I}(\sigma)-\ft12 \rmi \gamma^I_{JK}(\sigma)\bar{\psi}^{iJ}\psi^{jK}\; ,
\end{equation}
where $\gamma^I_{JK}$ is symmetric in its lower indices. The first term will turn out to be the on-shell counterpart of a Fayet-Iliopoulos term. By using the rules (\ref{transfsigma2}-\ref{transfpsi2}) we can calculate the transformation of $A^{(ij)I}$, which now only is compatible with (\ref{transfAij}) if the following equations hold:
\begin{eqnarray}\label{FI1}
&&\partial_Jf^{(ij)I}+2\gamma^I_{JK}f^{(ij)K}=0\; ,\\
&&\gamma^I_{LM}\gamma^M_{JK}=-\ft12 \partial_L \gamma^I_{JK}\; .\label{gamma1}
\end{eqnarray}
Note that the requirement (\ref{gamma1}) follows from the terms in $\delta_Q A^{(ij)I}$ which are cubic in the gaugino.

From the study of the symmetry algebra, we can infer two more conditions which should hold in the non-Abelian sectors of the gauge theory. As can be checked on the bosons, the commutator of a supersymmetry and a gauge transformation should vanish. On the fermions, this condition implies that $A^{(ij)I}$ transforms in the adjoint representation of the gauge group. This leads to the other defining conditions for the geometry:
\begin{eqnarray}\label{FI2}
&&f_{JL}{}^K\sigma^L\partial_Kf^{ijI}-f_{JK}{}^If^{Kij}=0\; ,\\
&&2f_{J(L}{}^M\gamma_{K)M}^I-f_{JM}{}^I\gamma_{KL}^M+f_{JM}{}^N\sigma^M\partial_N\gamma_{KL}^I=0\; .\label{gamma2}
\end{eqnarray}

We can now completely determine all equations of motion as the non-closure functional $\Gamma^{iI}$ transforms under Susy as
\begin{eqnarray}
\delta\Gamma^{iI}&=&-\ft38\rmi\gamma^I_{JK}\bar{\psi}^{iJ}\Gamma^{jK}\epsilon_j-\ft38\rmi\gamma^I_{JK}\bar{\psi}^{iJ}\gamma^\mu \Gamma^{jK}\gamma_\mu \epsilon_j+\ft1{16}\rmi\gamma^I_{JK}\bar{\psi}^{iJ}\gamma^{\mu \nu}\Gamma^{jK}\gamma_{\mu \nu}\epsilon_j\nonumber\\&&
-\ft12\rmi \Delta^I\epsilon^i-\ft12\Xi_\mu^I\gamma^\mu \epsilon^i\; ,
\end{eqnarray}
where $\Delta^I$ is the equation of motion for the real scalar $\sigma^I$ and $\Xi^I_\mu$ the one for the real vector $A_\mu^I$. All dynamical constraints thus read
\begin{eqnarray}
\Gamma^{iI}&=&\slashed{\mathfrak{D}}\psi^{iI}+\gamma^I_{JK}\slashed{\mathfrak{D}}\sigma^J\psi^{iK} +\ft12\rmi\gamma^I_{JK}\slashed{F}^J\psi^{iK}-\ft12 \partial_K\gamma_{JL}^I \bar{\psi}^{iJ}\psi^{jL}\psi^K_j+2\rmi g\gamma^I_{JK}f^{ijI}\psi_j^K \nonumber\\&&+\rmi gf_{JK}{}^I\sigma^J\psi^{iK}\equiv0\; ,\label{psiEOM}\\
\Delta^I&=&\Box \sigma^I+ \gamma_{JK}^I\mathfrak{D}_\mu \sigma^J\mathfrak{D}^\mu \sigma^K-\ft12 \gamma^I_{JK}F_{\mu \nu}^JF^{\mu \nu K}-\ft12 \partial_L\gamma^I_{JK}\bar{\psi}^L\slashed{\mathfrak{D}}\sigma^J\psi^K-\ft14\rmi\partial_K\gamma^I_{JL}\bar{\psi}^J\slashed{F}^L\psi^K\nonumber\\&&-\ft5{32} \partial_M\partial_K\gamma^I_{JL}\bar{\psi}^{Lj}\psi^{Jk}\bar{\psi}^M_k\psi_j^K
-\ft18\partial_K\gamma^I_{JL}\gamma^J_{MN}\bar{\psi}^{Kj}\psi^{Lk}\bar{\psi}^M_k\psi^N_j+\ft14\partial_K\gamma^I_{JL}\gamma^K_{MN}\bar{\psi}^{Lj}\psi^{Jk}\bar{\psi}^N_k\psi^M_j\nonumber\\&&
+\ft12\rmi g f_{JK}{}^I\bar{\psi}^J\psi^K+\rmi g\gamma^I_{JK}f^J_{LM}\sigma^M\bar{\psi}^L\psi^K
+\rmi g\partial_J\gamma^I_{LM}f^{ijL}\bar{\psi}_i^J\psi_j^M+\rmi g \partial_J\gamma^I_{LM}f^{ijJ}\bar{\psi}_i^L\psi_j^M\nonumber\\&&
+2g^2\gamma^I_{JK}f^{ijJ}f_{ij}^K\equiv0\; ,\label{sigmaEOM}\\
\Xi_\mu^I&=&\mathfrak{D}^\nu F_{\nu \mu}^I-\ft14\gamma^I_{JK}\varepsilon_{\mu \nu \rho \sigma \tau}F^{\nu\rho J} F^{\sigma \tau K} +2 \gamma^I_{JK}\mathfrak{D}^\nu \sigma^KF_{\nu \mu}^J+\rmi\gamma^I_{JK}\bar{\psi}^J\mathfrak{D}_\mu \psi^K\nonumber\\&&+\ft14 \partial_M\gamma_{JK}^I\bar{\psi}^M\gamma_\mu \slashed{F}^J\psi^K
-\ft12\rmi\partial_L\gamma_{JK}^I\bar{\psi}^L\gamma_\mu \slashed{\mathfrak{D}}\sigma^J\psi^K
-\ft5{32}\rmi\partial_M\partial_K\gamma^I_{JL}\bar{\psi}^{Lj}\psi^{Jk}\bar{\psi}^M_k\gamma_\mu \psi_j^K
\nonumber\\&&
-\ft18 \rmi \partial_{(K}\gamma_{L)J}^I\gamma^J_{MN}\bar{\psi}^{Mk}\psi^{Nj}\bar{\psi}^L_k\gamma_\mu \psi_j^K
-gf_{JK}{}^I\sigma^J\mathfrak{D}_\mu \sigma^K+\ft12gf_{JK}{}^I\bar{\psi}^J\gamma_\mu \psi^K\nonumber\\&&-g\partial_J\gamma^I_{LM}f^{ijL}\bar{\psi}^J_i\gamma_\mu\psi_j^M\equiv0\; .\label{AEOM}
\end{eqnarray}
Note that the last term in (\ref{sigmaEOM}) is a more general potential than in the off-shell case, since it can be present in the non-Abelian sectors of the theory.
As we already mentioned, the action, and hence the equations of motion for off-shell vector multiplets, are completely fixed by a tensor $C_{IJK}$. Similarly, the dynamical constraints in the on-shell case are determined by a new object $\gamma^I_{JK}$ which is symmetric in its lower indices. In the Abelian factors of the gauge theory and in the absence of a Fayet-Iliopoulos term, (\ref{gamma1}) is the only constraint and we will show in the next Section that it is the counterpart of the fact that $C_{IJK}$ is constant. In the non-Abelian case, the transformation of the object $\gamma_{JK}^I$ should be compatible with (\ref{gamma2}), which is to be compared to the demand that $C_{IJK}$ is gauge invariant. 

In the Abelian off-shell case, we can add a constant term (the Fayet-Iliopoulos term) to the equation of motion of the auxiliary field, which yields a potential in the action. The on-shell counterpart is the object $f^{(ij)I}$ which in the Abelian case satisfies (\ref{FI1}). In the non-Abelian case, this FI term should also satisfy (\ref{FI2}) but is not eliminated by it. This is a major generalisation as in the off-shell case non-Abelian FI terms are not possible.\footnote{In the case of local (off-shell) supersymmetry, $\mathfrak{su}(2)$ FI terms exist.}

\section{Action in Five Dimensions} \label{action}
In this Section, we will show that the existence of an action reduces the set of equations in Section \ref{5d} to the well-known ones of very special geometry. More specifically, we will show that we can construct a standard action only if the object $\gamma^I_{JK}$ can be linked to the above mentioned tensor $C_{IJK}$.

In a general non-linear sigma model, the scalars are considered to be coordinates on the target space. In a suitable formulation, the equations of motion then transform covariantly under coordinate transformations and as a consequence, there appears a connection in the kinetic term for the scalars  
\begin{equation}\label{defboxsc}
\Box \sigma^x=\partial_\mu\partial^\mu \sigma^x+\Gamma^x_{yz}\partial_\mu\sigma^y\partial^\mu\sigma^z\; .
\end{equation}
In general, this (torsionless) connection is affine. If we demand that this equation of motion is to be derivable from an action with a standard kinetic term, i.e.
\begin{equation}\label{defkinsc}
\mathcal{L}=-\ft12g_{xy}\partial_\mu \sigma^x\partial^\mu \sigma^y+\dots\; ,
\end{equation}
the connection is metric preserving. Thus, since the existence of an action requires the introduction of a new object, namely the metric on the target space, the target space geometry becomes Riemannian.   

This is a rather general observation for non-linear sigma models, but in our case more care might be needed. The reason is mainly that we are working in special coordinates in which target space diffeomorphisms are not manifest. Therefore, the equations of motion (\ref{psiEOM})-(\ref{AEOM}) do not seem to be covariant under general coordinate transformations and the object $\gamma^I_{JK}$ should rather be seen as the on-shell counterpart of $C_{IJK}$ than to be considered a connection. This cautionary remark aside, we will proof that in the off-shell case $\gamma^I_{JK}$ can be related to the Levi-Civita connection corresponding to the special real metric (in special coordinates) $g_{IJ}=C_{IJK}\sigma^K$. 

Without referring yet to the explicit form of the Lagrangian, we need one more object to characterise the theory if the equations of motion (\ref{psiEOM})-(\ref{AEOM}) should be derivable from an action functional $S$, as we have
\begin{equation}\label{EOMpsi}
\frac{\delta S}{\delta \bar{\psi}_i^I}=g_{IJ}\Gamma^{iJ}\; ,
\end{equation}
and similarly for the other equations of motion.
We will suppose that the object $g_{IJ}$ depends on the scalars $\sigma^I$ only.  

Paralleling the discussion in \cite{Stelle:2003rr}, we will use the Batalin-Vilkovisky formalism \cite{Batalin:1983jr,Henneaux:1990jq,Gomis:1995he} to retrieve the conditions for the existence of an action. The main advantage of this formalism is that the explicit form of the classical action should not be known, although we will list it at the end of the Section. In the BV formalism we introduce a ghost for every symmetry. If the symmetry is rigid, the ghost field is taken to be constant. Moreover, for every field, we introduce an anti-field. The BV action $S_{BV}$ now keeps track of all algebraic relations between the fields by ensuring the validity of the master field equation 
\begin{equation}\label{master}
(S_{BV},S_{BV})=0\; .
\end{equation}
We will now introduce translational ghosts $c^\mu$, supersymmetry ghosts $c^i$, gauge symmetry ghosts $\alpha^I$ and anti-fields, and expand the BV action up to anti-field number 2.
\begin{eqnarray}
S_{BV}&=&\int d^5x\; \mathcal{L}_0+\mathcal{L}_1+\mathcal{L}_2+\dots\; ,\nonumber\\
\mathcal{L}_1&=&\sigma_I^*c^\mu\partial_\mu\sigma^I+A_I^{*\mu}c^\nu\partial_\nu A_\mu^I +\bar{\psi}_{iI}^*c^\mu\partial_\mu\psi^{iI}+\ft12\rmi\sigma_I^*\bar{\psi}^Ic +\ft12A_I^{*\mu}\bar{\psi}^I\gamma_\mu c\nonumber\\&& 
+\bar{\psi}^*_{iI}\left(-\ft12 \rmi \slashed{\mathfrak{D}}\sigma^Ic^i-\ft14\slashed{F}^Ic^i-\ft12\rmi\gamma^I_{JK}\bar{\psi}^{iJ}\psi^{jK}c_j +f^{(ij)I}c_j\right)-g\sigma_I^*f^I{}_{JK}\alpha^J\sigma^K\nonumber\\&&
+A_I^{\mu *}\mathfrak{D}_\mu \alpha^I-g\bar{\psi}^*_{iI}f^I{}_{JK}\alpha^J\psi^{iK}\; ,\nonumber\\
\mathcal{L}_2&=& -\ft14 c_\mu^*\bar{c}\gamma^\mu c+\bar{c}^{(i}\psi_I^{*j)}\bar{c}_i\psi_{jJ}^*g^{IJ}
-\ft12g\alpha_I^*f^I{}_{JK}\alpha^J\alpha^K+\ft14\rmi \alpha_I^*\sigma^I\bar{c}c+\ft14\alpha_I^*A_\mu^I\bar{c}\gamma^\mu c\;, 
\end{eqnarray}
where $\mathcal{L}_0$ denotes the classical (unknown) Lagrangian. Of course, the BV action can have terms with higher anti-field number and to consider the form of these terms, we need to know the dimension, ghost number and anti-field number of each field (see Table \ref{dimtable}). For consistency the dimension of each field together with its anti-field should add up to the same number, which we take to be equal to one. As a consequence, none of the fields have negative dimension. Using the fact that each term in the Lagrangian should have dimension two and vanishing ghost number, we can construct terms with higher anti-field number, but none of these terms will spoil the arguments below. 

\begin{table}[!tb]
\begin{center}
\begin{tabular}{|c|c|c|c||c|c|c|c|}
\hline
field&dim&gh&afn&anti-field&dim&gh&afn\\
\hline \hline
$\sigma^I$&$0$&$0$&$0$&$\sigma_I^*$&$1$&$-1$&$1$\\
$A_\mu^I$&$0$&$0$&$0$&$A_I^{*\mu}$&$1$&$-1$&$1$\\
$\psi^{iI}$&$1/2$&$0$&$0$&$\psi_{iI}^*$&$1/2$&$-1$&$1$\\
$c^\mu$&$0$&$1$&$0$&$c_\mu^*$&$1$&$-2$&$2$\\
$c^i$&$1/2$&$1$&$0$&$c_i^*$&$1/2$&$-2$&$2$\\
$\alpha^I$&$0$&$1$&$0$&$\alpha_I^*$&$1$&$-2$&$2$\\
$\mathcal{L}$&$2$&$0$&$-$&&&&\\
\hline
\end{tabular}
\end{center}
\caption{Dimension, ghost number and anti-field number of all fields}\label{dimtable}
\end{table}

In the master equation (\ref{master}) we concentrate on terms cubic in the supersymmetry ghost $c^I$ and quadratic in the gaugino anti-field $\psi_{iI}^*$ and find
\begin{eqnarray}
&&\ft12 \rmi \partial_Ig^{JK}\bar{\psi}^Ic \bar{c}^{(i}\psi^{*j)}_J\bar{c}_i\psi_{jK}^*-2\rmi g^{IJ}\gamma^K_{IL}\bar{\psi}^*_{(iK}c_{k)}\bar{\psi}^{kL}c_j\bar{c}^{(i}\psi^{*j)}_J\nonumber\\&&=
\ft12\rmi\bar{\psi}^Ic \bar{c}^{(i}\psi^{* j)}_J\bar{c}_i\psi^*_{jK}(\partial_Ig^{JK}+2\gamma^{(J}_{IL}g^{K)L})
+2\rmi\gamma_{IL}^{[J}g^{K]L}\bar{c}_{(i}\psi_{j)J}^*\bar{c}^{(i}\psi^{*k)}_K\bar{\psi}^{jI}c_k=0\; .
\end{eqnarray}
This is equivalent with the following conditions. 
\begin{eqnarray}
&&\partial_Ig^{JK}+2g^{L(J}\gamma^{K)}_{IL}=0\; ,\label{covconst}\\
&&\gamma_{I,JK}=g_{IL}\gamma^L_{JK}=\gamma_{(I,JK)}\; .\label{1christ=C}
\end{eqnarray}
The first condition means that the object $\gamma$ introduced in the previous Section can now be seen as a Levi-Civita connection with respect to the metric $g_{IJ}$. The second is the complete symmetry of the first Christoffel connection coefficients. Applying those conditions to (\ref{gamma1}), we find that these first Christoffel connection coefficients should be constant, and we will take
\begin{equation}\label{defC}
\gamma_{I,JK}=\ft12 C_{IJK}\; .
\end{equation}
This relation implies that $g_{IJ}=C_{IJK}\sigma^K+a_{IJ}$ where the last term is an integration constant. 
In the Abelian case, the only other condition we need to consider is (\ref{FI1}), which implies that $f^{(ij)}_I=g_{IJ}f^{(ij)J}$ is constant.

The non-Abelian case is slightly more involved as there are two more conditions to be solved. Using the above expressions for the metric and the connection, (\ref{gamma2}) can first of all be solved trivially if the metric is constant, as the corresponding $\gamma^I_{JK}$ is zero. If the three-tensor $C_{IJK}$ is differing from zero, the same condition (\ref{gamma2}) is solved if $a_{IJ}=0$ and the three-tensor is gauge-invariant
\begin{equation}
f_{I(J}{}^MC_{KL)M}=0\; .
\end{equation}
In both cases, the condition on the FI terms (\ref{FI2}) implies that they are vanishing.

To conclude this Section, we give the action (for $a_{IJ}=0$) together with the field equations.  
\begin{eqnarray}
S&=&\int d^5x \; C_{IJK}\sigma^K(-\ft12\mathfrak{D}_\mu \sigma^I\mathfrak{D}^\mu \sigma^J-\ft14F^I_{\mu \nu}F^{J\mu \nu}-\ft12\bar{\psi}^I\slashed{\mathfrak{D}}\psi^J)-g^2C_{IJK}\sigma^If^J_{ij}f^{ijK}\nonumber\\&&
-\ft18\rmi C_{IJK}\bar{\psi}^I\slashed{F}^J\psi^K-\ft14\rmi g f_{IJ}{}^KC_{KLM}\sigma^L\sigma^M\bar{\psi}^I\psi^J
+\ft12\rmi gC_{IJK}f^{ijI}\bar{\psi}_i^J\psi_j^K\nonumber\\&&
 +\ft1{16} C_{IJM}C_{KLN}g^{MN}\bar{\psi}^{iI}\psi^{jJ}\bar{\psi}_i^K\psi_j^L\nonumber\\&&
-\ft1{24}C_{IJK}\varepsilon^{\mu \nu \rho \sigma \tau}A_\mu^I(F^J_{\nu \rho}F^K_{\sigma \tau} +gf_{LM}{}^JA_\nu^LA_\rho^MF^K_{\sigma \tau}+\ft25g^2f_{LM}^Jf_{NP}{}^KA_\nu^LA_\rho^MA_\sigma^NA_\tau^P)\; ,\nonumber\\
\frac{\delta S}{\delta \bar{\psi}_i^I}&=&g_{IJ}\Gamma^{iJ}\; ,\quad 
\frac{\delta S}{\delta \sigma^I}=g_{IJ}\Delta^J-\ft12 C_{IJK}\bar{\psi}^J\Gamma^K\; ,\quad  
\frac{\delta S}{\delta A^{\mu I}}=g_{IJ}\Xi_\mu^J-\ft12 \rmi C_{IJK}\bar{\psi}^J\gamma^\mu \Gamma^K\; .\nonumber\\
\end{eqnarray}
The action without FI terms equals the one in \cite{Bergshoeff:2002qk} after elimination of the auxiliary field $Y^{ijI}$ by its algebraic equation of motion. 
\section{On-Shell Vector Multiplet in Four Dimensions}\label{4d}
We can now parallel the construction of Section \ref{5d} to give a generalisation of special geometry. The on-shell field content of a vector multiplet is given by a complex scalar $X$, a vector $A_\mu$ and two fermions $\Omega_i$ and $\Omega^i$ \cite{Sierra:1983cc,Gates:1984py}. As in the five-dimensional case, the off-shell multiplet has in addition an auxiliary real scalar field in the $\mathbf{3}$ of the $\mathfrak{su}(2)$ R-symmetry group. In the present context, the field-dependent object $A^{(ij)I}$ in (\ref{transfomega}) stands in its place. 

The supersymmetry transformation rules again are very similar to the off-shell case.
\begin{eqnarray}
\delta X^I&=&\bar{\epsilon}^i\Omega_i^I-gf_{JK}{}^I\alpha^J X^K\; ,\\
\delta A_\mu^I&=& \varepsilon^{ij}\bar{\epsilon}_i\gamma_\mu \Omega_j^I+\mbox{h.c.}+\partial_\mu \alpha^I-gf_{JK}{}^I \alpha^J A_\mu^K\; ,\\
\delta \Omega_i^I&=&\slashed{\mathfrak{D}}X^I\epsilon_i+\ft14 \slashed{F}^I \varepsilon_{ij}\epsilon^j-A_{(ij)}^I\epsilon^j+gf_{JK}{}^IX^J\bar{X}^K\varepsilon_{ij}\epsilon^j-gf_{JK}{}^I\alpha^J \Omega_i^K\; .\label{transfomega}
\end{eqnarray}     
In our conventions (see also the Appendix), the position of the $\mathfrak{su}(2)$ indices on the spinors denote their chirality.
The field-dependent tensor $A_{(ij)}^I$ should transform as 
\begin{equation}\label{transfA4} 
\delta A_{(ij)}^I=\bar{\zeta}_{(i|I}\epsilon_{|j)}-\bar{\zeta}^{k I}\epsilon^l \varepsilon_{k(i}\varepsilon_{j)l}\; .
\end{equation}
The commutator of two supersymmetries now again yields a translation, a field dependent gauge transformation and non-closure functionals (when imposed on the fermions), which are interpreted as dynamical constraints on the physical field $\Omega_i^I$, hence as equations of motion. 
\begin{eqnarray}
[\delta_1,\delta_2]\Omega_i^I&=&\xi^a\mathfrak{D}_a\Omega_i^I+\delta_G(2\varepsilon^{ij}\bar{\epsilon}_{2i}\epsilon_{1j}X+{\rm h.c.})\Omega_i^I-\ft14 \xi^a\gamma_a\Gamma_i^I-\ft12\bar{\epsilon}_{i[2}\gamma^\mu\epsilon_{1]}^j\gamma_\mu \Gamma_j^I\nonumber\\&&-\ft34 \varepsilon_{ji}\varepsilon_{lk}\bar{\epsilon}_1^k\epsilon_2^l\Gamma^{jI}+\ft18 \varepsilon_{ji}\varepsilon_{lk}\epsilon_1^{(j}\gamma_{ab}\epsilon^{l)}_2\gamma^{ab}\Gamma^{kI}\; ,
\end{eqnarray} 
with $\xi^a=\bar{\epsilon}_2^i\gamma^a\epsilon_{1i}+\bar{\epsilon}_{2i}\gamma^a\epsilon_1^i$. We now choose an Ansatz similar to the five-dimensional case
\begin{equation}\label{AnsatzAijd4}
A_{(ij)}^I=-\ft12\gamma^I_{JK}\bar{\Omega}_{(i}^J\Omega_{j)}^K+\ft12\bar{\gamma}^I_{JK}\varepsilon_{k(i}\varepsilon_{j)l} \bar{\Omega}^{kJ}\Omega^{lK}\; ,
\end{equation}
where the object $\gamma^I_{JK}$ both depends on the complex scalar $X$ and its complex conjugate.
Note that for simplicity, we neglected the possibility of adding a Fayet-Iliopoulos term.

Determining the defining conditions for the geometry is a little bit more subtle than in the five-dimensional case.   
Again, we have to demand that the Ansatz (\ref{AnsatzAijd4}) is compatible with (\ref{transfA4}). The defining conditions then follow from considering the terms cubic in the fermionic fields. If all fields have the same chirality, the part completely symmetric in the gauge indices assumes already the correct form (i.e. is compatible with (\ref{transfA4})), due to (\ref{Symm3}). E.g.
\begin{equation}
\bar{\epsilon}^k\Omega_{(i}^{(I}\bar{\Omega}_{j)}^J\Omega_k^{K)}= \ft12\varepsilon_{k(i}\bar{\Omega}_{j)}^{(I}\Omega_l^{J|}\bar{\epsilon}^k\Omega_{m}^{|K)}\varepsilon^{lm}\; .
\end{equation}
Therefore, the first equality of (\ref{defgeom1d4}) will  only restrict the part anti-symmetric in $JL$.
In summary, the defining conditions for the geometry now read
\begin{equation}\label{defgeom1d4}
\gamma_{K[J}^I\gamma_{L]M}^K=-\partial_{[J}\gamma_{L]M}^I\; , \quad \gamma_{JK}^I\bar{\gamma}_{LM}^K=-\partial_J\bar{\gamma}_{LM}^I\; ,\quad \mbox{h.c.}
\end{equation}
As in the five-dimensional case, the symmetry algebra imposes another restriction in the non-Abelian case, as the tensor $A_{(ij)}^I$ should transform in the adjoint under gauge transformations, which implies a condition similar to (\ref{gamma2}).
\begin{equation}
\partial_M\gamma^I_{JK}f^M_{LN}X^N+\bar{\partial}_{M}\gamma^I_{JK}f^M_{LN}\bar{X}^N-f^I_{LM}\gamma^M_{JK}+2f^M_{L(J}\gamma^I_{K)M}=0\; .\label{defgeom2d4}
\end{equation}
With all this information, it is possible to find the complete non-closure functional for the fermions, which transforms into the other equations of motion.
\begin{eqnarray}\label{gamma4d}
\Gamma_i^I&=&\slashed{\mathfrak{D}}\Omega_i^I+\gamma^I_{JK}\slashed{\mathfrak{D}}X^J\Omega^K_i +\ft14\bar{\gamma}^I_{JK}\slashed{F}^J\Omega^{Kj}\varepsilon_{ji}-2gf^I_{JK}X^K\Omega^{jJ}\varepsilon_{ji} +g\bar{\gamma}^I_{JK}f^K_{LM}X^L\bar{X}^M\Omega^{jJ}\varepsilon_{ji}\nonumber\\&&
-\ft12 \partial_{[\bar{L}}\bar{\gamma}^I_{J]K}\bar{\Omega}^{lJ}\Omega^{jK}\Omega^{kL}\varepsilon_{lk}\varepsilon_{ji}
-\ft14\partial_{(\bar{L}}\bar{\gamma}_{J)K}^I\varepsilon_{ji}\varepsilon_{kl}\bar{\Omega}^{kJ}\Omega^{jK}\Omega^{lL}
+\ft14\bar{\gamma}_{M(L}^I\gamma^M_{J)K}\varepsilon_{ji}\varepsilon_{kl}\bar{\Omega}^{kJ}\Omega^{jK}\Omega^{lL}\nonumber\\&&
-\ft12 \partial_{\bar{L}}\gamma^I_{JK}\bar{\Omega}^J_j\Omega_i^K\Omega^{jL}\; ,\\ \label{delta4d}
\Delta^I&=&\Box X^I+\gamma^I_{JK}\mathfrak{D}_\mu X^J\mathfrak{D}^\mu X^K 
-\ft18 \bar{\gamma}^I_{JK}F^J_{\mu \nu}F^{K\mu \nu}-\ft1{16}\rmi\varepsilon_{\mu \nu \rho \sigma}\bar{\gamma}^I_{JK}F^{J\mu \nu}F^{K\rho \sigma}\nonumber\\&&
+2g^2f^I_{JK}f^J_{LM}X^KX^L\bar{X}^M-g^2\bar{\gamma}^I_{JK}f^K_{LM}f^J_{NP}X^L\bar{X}^MX^N\bar{X}^P+\mbox{fermions}\; , \\\label{xi4d}
\Xi_\mu^I&=&\mathfrak{D}^\nu F_{\nu \mu}^I+\Big(\gamma^I_{JK}\mathfrak{D}^\nu X^JF_{\nu \mu}^K+\ft12\rmi\varepsilon_{\mu \nu \rho \sigma}\gamma^I_{JK}\mathfrak{D}^\nu X^JF^{\rho \sigma K}\nonumber\\&&
+2gf_{JK}{}^I\mathfrak{D}_\mu X^J\bar{X}^K+2g\gamma^I_{JK}f_{LM}{}^KX^L \bar{X}^M\mathfrak{D}_\mu X^J+\mbox{h.c.}\Big) +\mbox{fermions}\; . 
\end{eqnarray}
As in the five-dimensional case, the target-space geometry is specified by an object $\gamma^I_{JK}$ which depends on the scalar fields and for which the conditions (\ref{defgeom1d4}) and (\ref{defgeom2d4}) hold. 
\section{Action in Four Dimensions}\label{4daction}
As in five dimensions, the introduction of an action requires the existence of a metric $g$ on the target space.  
Rather than repeating the complete discussion in Section \ref{action}, we will now immediately list the conditions imposed on this metric as they are found using the BV formalism. First of all, the metric turns out to be hermitian with respect to the standard complex structure $J=\rmi dX^I\otimes \partial_I-\rmi d\bar X^I\otimes \bar \partial_I$. The other requirements are
\begin{eqnarray}
&&\partial_Ig^{J\bar K}+\gamma_{IL}^Jg^{L\bar K}=0\; , \label{hermcon}\\
&&g_{I\bar J}=\overline{g_{I\bar J}}=g_{\bar I J}\label{realg}\; ,\\
&&g^{\bar L[I}\gamma^{J]}_{KL}=0\; .\label{complsym}
\end{eqnarray}
The first relation (\ref{hermcon}) states that $\gamma^I_{JK}$ is the Levi-Civita connection with respect to this hermitian metric, while the next equation (\ref{realg}) states that the metric is real. The last condition (\ref{complsym}) again implies the complete symmetry of the first Christoffel connection coefficients, i.e.
\begin{equation}
g_{\bar I L}\gamma^L_{JK}=\gamma_{\bar I,JK}=\gamma_{(IJK)}\; .
\end{equation}
Using the above in the defining equations (\ref{defgeom1d4}), it is easy to prove that this Christoffel coefficients are derivable from a holomorphic functional $F(X)$ of the scalar fields $X$. The same functional can be used to specify the metric.
\begin{eqnarray}
&&\gamma_{\bar I,JK}=-2\rmi F_{IJK}\; ,\label{deflc4d} \\
&&g_{I\bar J}=-2\rmi (F_{IJ}-\bar{F}_{IJ})\label{defg4d}\; .
\end{eqnarray}
For this, we adapted the following notation.
\begin{eqnarray}\label{defFI}
F_I(X)&=&\frac{\delta F(X)}{\delta X^I}\; , \quad F_{IJ}(X)=\frac{\delta^2 F(X)}{\delta X^I\delta X^J}\; , \quad \dots\nonumber\\
\bar F_I(X)&=&\frac{\delta \bar F(X)}{\delta \bar X^I}\; , \quad \bar F_{IJ}(X)=\frac{\delta^2 \bar F(X)}{\delta \bar X^I\delta \bar X^J}\; , \quad \dots
\end{eqnarray}

To discuss the other condition (\ref{defgeom2d4}) which only holds for non-Abelian theories, we need to know how the holomorphic functional $F(X)$ transforms under gauge transformations. To be as general as possible, we take
\begin{equation}
\delta_G F=-g\alpha^I F_J f^J{}_{IK}X^K=-g\alpha^I C_I\; ,
\end{equation}
where $C_I$ is a general holomorphic functional of the scalar fields $X$. Using (\ref{defg4d}) and (\ref{deflc4d}) we can rewrite (\ref{defgeom2d4}) as
\begin{equation}
\partial_I\partial_J\partial_KC_L+2\rmi g^{N\bar P}F_{JKP}(\partial_N\partial_IC_L-\partial_{\bar{N}}\partial_{\bar{I}}\bar{C}_{\bar{L}})=0\; .
\end{equation}
Note that the two known cases \cite{deWit:1985px,deWit:1987ph} $\delta_G F=0$ and $\delta_G F=-g\alpha^I C_{I,JK}X^JX^K$ with $C_{I,JK}$ real, are certainly solutions to the above equation. 
 
In conclusion, we have shown that if the equations of motion (\ref{gamma4d})-(\ref{xi4d}) can be deduced from an action, $\gamma^I_{JK}$ is the Levi-Civita connection in special coordinates and the theory is completely specified by a holomorphic functional of the scalar fields. 
\section{Dimensional Reduction}\label{dimred}
The five-dimensional vector multiplet can be considered to be the dimensional reduction of the $\mathcal{N}=1$ vector multiplet in six dimensions. This multiplet does not contain scalars and the five-dimensional real scalar field originates from the sixth component of the higher dimensional vector. If one further dimensionally reduces to four dimensions, one arrives at a theory with a complex scalar as there is again a component of the five-dimensional vector which transforms trivially under the broken Poincar\'e group.

The reduction of off-shell vector multiplets from five to four dimensions implies a mapping of a very special real manifold to a complex one, called a very special K\"ahler manifold. The map is called the $\mathbf{r}$-map \cite{deWit:1992nm}. Such target spaces of four-dimensional vector multiplets are not the most general ones, as the process of dimensional reduction leads to isometries in the target space. Therefore, the image of very special real manifolds under the $\mathbf{r}$-map is a subset of all special K\"ahler manifolds. A similar consideration holds in the present context. 

We will now perform the Kaluza-Klein reduction of the theory presented in Section \ref{5d} (without FI-like terms) to the one given in Section \ref{4d}. Therefore, we will suppose that the fifth dimension is a circle, and that all fields are periodic around this circle. In the Fourier expansion of these fields we will only retain the zero mode. Moreover, the five-dimensional Poincar\'e group is broken, such that the fifth direction cannot be rotated into the others. As for the explicit Ansatz, the first four gamma matrices remain, the fifth becomes the chirality operator $\hat{\gamma}_4=\gamma_*$. A five-dimensional spinor reduces to a left- and a right-handed four-dimensional one. More exactly, the reduction Ansatz for the fermions, the Susy parameter and the non-closure functionals are
\begin{equation}
\hat{\psi}_i^I=\sqrt{2}(\Omega_i-\Omega^j\varepsilon_{ji})\; , \quad
\hat{\epsilon}_i=\sqrt{2}(\epsilon_i+\epsilon^j\varepsilon_{ji})\;,\quad
\hat{\Gamma}_i^I=\sqrt{2}(\Gamma_i^I-\Gamma^{jI}\varepsilon_{ji})  \;  . 
\end{equation}
The first four components of the vector remain. The fifth component $\hat{A}_4^I$ together with the real scalar $\hat{\sigma}^I$ form the complex scalar field $X^I$. The dimensional reduction of the bosonic equations of motion works similarly. 
\begin{equation}
X^I=\ft12(\hat{A}_4^I-\rmi \hat{\sigma}^I)\; ,\quad \Delta^I=\ft12 (\hat{\Xi}_4^I-\rmi \hat{\Delta}^I)\;,\quad \hat{\Xi}_\mu^I=\Xi_\mu^I\; .
\end{equation}
As a consequence of the fact that $\hat{A}_{(ij)}^I=A_{(ij)}^I$,\footnote{This can easily be seen from the dimensional reduction of the supersymmetry transformation rules.} we find that
\begin{equation}\label{redgamma}
\hat{\gamma}^I_{JK}=-\ft12\rmi\gamma^I_{JK}=+\ft12\rmi\bar{\gamma}^I_{JK}\; , 
\end{equation}
which means that the reduced $\gamma_{JK}^I$ is purely imaginary. Due to (\ref{redgamma}), the four-dimensional $\gamma$'s only depend on the five-dimensional scalar, i.e.
\begin{equation}
\hat \gamma^I_{JK} (\sigma) \to \gamma^I_{JK}(X-\bar{X})\; .
\end{equation}
The reduction of (\ref{gamma1}) then yields
\begin{equation}\label{redgamma1}
\partial_L\gamma^I_{JK}=-\gamma^I_{LM}\gamma^M_{JK}\; , \quad \partial_{\bar L} \gamma^I_{JK}=-\bar\gamma^I_{LM}\gamma^M_{JK}\; ,
\end{equation}
which is a stricter demand than (\ref{defgeom1d4}), while the reduction of (\ref{defgeom2d4}) does not render any new information. 

In the off-shell case, the theories in the image of the $\mathbf{r}$-map have holomorphic functionals of the form $F(X)=C_{IJK}X^IX^JX^K$, which is only a subset of the allowed functionals. Similarly for on-shell vector multiplets, dimensionally reduced $\gamma^I_{JK}$'s satisfy (\ref{redgamma1}) and hence are more restricted than in generic four-dimensional theories. 
\section{Conclusions}\label{conclusions}
In this letter we constructed on-shell rigid supersymmetric $\mathcal{N}=2$ vector multiplets in five and in four dimensions. The equations of motion were found as dynamical constraints on the fields imposed by the closure of the supersymmetry algebra. It was found that the target-space geometry is specified by an object $\gamma^I_{JK}$ which generalises the results from (very) special geometry. Due to new Fayet-Iliopoulos terms, we were able to find more general potentials than the on-shell non-Abelian theory allows for. It was also discussed that the requirement of an action restricts the theories to the well-known ones of (very) special geometry. A generalisation of the $\mathbf{r}$-map was also given by considering dimensional reduction. 

There are various possible ways of further study. First of all, the meaning of the $\gamma^I_{JK}$'s might be clarified. One could also use this multiplet to gauge isometries in hypercomplex manifolds. This construction might then be used to perform a hypercomplex quotient. One could also couple these multiplets to supergravity to find generalisations of local (very) special geometry.
\section*{Acknowledgements}
We would like to thank Toine Van Proeyen for stimulating discussions and careful reading of the manuscript. 
This work has been supported in part by the FWO-Vlaanderen, the European Community's Human Potential Programme under contract HPRN-CT-2000-00131 Quantum Spacetime and the Federal Office for Scientific, Technical and Cultural Affairs through the Inter-University Attraction Pole P5/27.
\appendix
\section{Conventions}
We use the conventions of \cite{VanProeyen:1999ni} but highlight some important aspects here. 
\subsection{Four Dimensions}
Anti-symmetrisation and symmetrisation, respectively denoted by $[\dots]$ and $(\dots)$, is done with unit weight.
The gamma matrices satisfy the following Clifford algebra.
\begin{equation}
\{\gamma_\mu,\gamma_\nu\}=2\eta_{\mu \nu}\; ,
\end{equation}
where the Minkowski metric is mostly plus. The chirality matrix is $\gamma_*=\rmi \gamma_0\gamma_1\gamma_2\gamma_3$. 
The duality conditions for the gamma matrices are
\begin{equation}
1=+\ft1{4!}\rmi \varepsilon_{abcd}\gamma_*\gamma^{abcd}\; , \quad
\gamma_a=-\ft1{3!}\rmi \varepsilon_{abcd}\gamma_*\gamma^{abc}\; , \quad
\gamma_{ab}=-\ft12\rmi \varepsilon_{abcd}\gamma_*\gamma^{cd}\; .
\end{equation}

The spinors are taken to be Majorana. The position of the $\mathfrak{su}(2)$ index indicates the chirality of the spinor. A left-handed spinor has eigenvalue $+1$ wrt the chirality matrix, and $a_\lambda$ denotes the chirality $\gamma_* \lambda=a_\lambda \lambda$. We used two Fierz identities frequently.
\begin{itemize}
\item For $a_\lambda=a_\psi$, we have
\begin{equation}\label{Fierzd4}
\lambda \bar{\zeta}\psi=-\ft12 \psi \bar{\zeta}\lambda+\ft18 \gamma_{ab}\psi \bar{\zeta}\gamma^{ab}\lambda\; .
\end{equation}
\item For $a_\lambda=-a_\psi$, we have
\begin{equation}
\lambda \bar{\zeta}\psi=-\ft12 \gamma_a\psi \bar{\zeta}\gamma^a\lambda\; .
\end{equation}
\end{itemize}
We define the number $t_{(n)}$ by 
\begin{equation}\label{eqsigns}
\bar{\lambda}^i \gamma_{(n)} \chi_i=t_{(n)} \bar{\chi}_i\gamma_{(n)} \lambda^i\; ,
\end{equation}
and its value can be found in Table \ref{signs}. If at least one of the spinors is commuting, the right-hand side in (\ref{eqsigns}) and in the above Fierz-identities should be multiplied by $-1$. This is important when using the Batalin-Vilkovisky formalism where the anti-field of a fermion is a commuting spinor, as is the ghost field corresponding to a supersymmetry.

From (\ref{Fierzd4}) and (\ref{eqsigns}), we can deduce an important Fierz-identity (which trivially holds for commuting spinors).
\begin{equation}\label{Symm3}
\lambda^{(i}\bar{\lambda}^j\lambda^{k)}=0\; .
\end{equation}
\subsection{Five Dimensions}
Most of the conventions are the same. Spinors are taken to be symplectic Majorana. The four-dimensional chirality matrix becomes $\gamma_4$. On a spinor, the $\mathfrak{su}(2)$ index has no additional meaning. Therefore it is not written in a bispinor that is a scalar under the R-symmetry group : $\bar \lambda^i \gamma^{(n)} \psi_i\equiv\bar \lambda \gamma^{(n)} \psi$. Raising and lowering of $\mathfrak{su}(2)$ indices is now done using the North-West South-East convention:
\begin{equation}
\varepsilon^{ij}A_j=A^i\; , \quad A^j\varepsilon_{ji}=A_i\; .
\end{equation}
The signs $t_{(n)}$ defined in (\ref{eqsigns}) are also shown in Table \ref{signs}. The basic Fierz identity is
\begin{equation}
\lambda_i \bar{\zeta}^i=-\ft14(\bar{\zeta}\lambda+\bar{\zeta}\gamma^a\lambda\gamma_a-\ft12\bar{\zeta}\gamma^{ab}\lambda\gamma_{ab})\; .
\end{equation}
For commuting spinors, the sign $t_{(n)}$ should be opposite and the right-hand side of the Fierz-identity should be multiplied by $-1$.
\begin{table}[!htb]
\begin{center}
\begin{tabular}{|c||c|c|c|c|}
\hline 
n=&0,4&1,5&2&3\\
\hline\hline
d=5&+&+&$-$&$-$\\
\hline
d=4&+&$-$&$-$&+\\
\hline
\end{tabular}
\end{center}
\caption{$t_{(n)}$ in various dimensions}\label{signs}
\end{table}
\providecommand{\href}[2]{#2}\begingroup\raggedright\endgroup

\end{document}